\documentclass[%
 aip,
 amsmath,amssymb,
 reprint,%
]{revtex4-1}

\usepackage{graphicx}
\usepackage{dcolumn}
\usepackage{bm}

\usepackage[utf8]{inputenc}
\usepackage[T1]{fontenc}
\usepackage{mathptmx}
\usepackage{etoolbox}
\usepackage{xcolor}
\usepackage{hyperref}
\makeatletter
\def\@email#1#2{%
 \endgroup
 \patchcmd{\titleblock@produce}
  {\frontmatter@RRAPformat}
  {\frontmatter@RRAPformat{\produce@RRAP{*#1\href{mailto:#2}{#2}}}\frontmatter@RRAPformat}
  {}{}
}%
\makeatother
\begin{document}

\preprint{AIP/123-QED}

\title[This article appeared in: \textbf{Phys. Fluids \textcolor{red}{36}, 033345 (2024);\href{https://doi.org/10.1063/5.0201053}{ \textcolor{blue}{doi: 10.1063/5.0201053}}}]{A smart granular intruder}

\author{Lázaro Martínez-Ortíz}
\affiliation{%
Group of Complex Systems and Statistical Physics, Physics Faculty, University of Havana, 10400 Havana, Cuba
}%

\author{Alex Rivera-Rivera}%
 
\affiliation{ 
Espoleta Tecnologías, S. R. L., 32 No. 119, Miramar, 11300 Havana, Cuba.
}%

\author{Ernesto Altshuler$^*$}
\email{ealtshuler@fisica.uh.cu}
\affiliation{%
Group of Complex Systems and Statistical Physics, Physics Faculty, University of Havana, 10400 Havana, Cuba
}%

\date{\today}

\begin{abstract}
It has been recently reported that irregular objects sink irregularly when released in a granular medium: a subtle lack of symmetry in the density or shape of a macroscopic object may produce a large tilting and deviation from the vertical path when released from the free surface of a granular bed. This can be inconvenient --even catastrophic-- in scenarios ranging from buildings to space rovers. Here, we take advantage of the high sensitivity of granular intruders to shape asymmetry: we introduce a granular intruder equipped with an inflatable bladder that protrudes from the intruder's surface as an autonomous response to an unwanted tilting. So, the intruder's symmetry is only slightly manipulated, resulting in the rectification of the undesired tilting. Our smart intruder is even able to rectify its settling path when perturbed by an external element, like a vertical wall. The general concept introduced here can be potentially expanded to real-life scenarios, such as ``smart foundations'' to mitigate the inclination of constructions on a partially fluidized soil.
\end{abstract}

\maketitle

\section{Introduction}
Granular matter --ensembles of objects ranging from grains of sand to asteroids-- can display fluid-like and solid-like features, frequently resulting in complex behaviours. 
For example, when a solid object sediments into a typical granular material –like sand– grains start flowing around the intruder in a liquid-like fashion, but the penetration eventually stops as the grains jam underneath, creating a solid phase.

The penetration of ``real-world objects'' in granular beds is usually affected by asymmetry --either inherent to the intruder or external to it-- causing undesired 
consequences in the sedimentation process. Dangerous tilting of building can be caused by ``soil liquefaction'' associated to earthquakes \cite{sancio2003ground}, 
while incorrect settling of submarine gravity-based structures, pipelines, armor blocks and anchors are feared by engineers \cite{kirca2019sinking} as 
much as the non-homogeneous settlement of gravel and pebbles in vibrated fresh concrete \cite{cai2021experimental}. 
In forensic science and archaeology, ancient objects --including bones-- can travel in nontrivial ways as they settle \cite{Mickleburgh2018actualistic}, 
which can eventually mislead archeological interpretation \cite{Moeyersons1977}. Even the technology involved in space exploration can be invalidated 
due to their uncontrolled sedimentation in a granular soil \cite{RoverSpirit2010}.

Understanding the physical nature of the settling of intruders in granular beds is mandatory to predict and mitigate the adverse scenarios described above. 
However, most research has concentrated on simple, symmetrical objects such as cylinders and spheres. In fact, a ``unified force law'' for granular penetration 
has been established by systematic experimentation with spherical intruders \cite{katsuragi2007unified}, and has been subsequently extended to take into 
account confinement effects \cite{pacheco2011infinite}. These laws have been tested even for gravities other than Earth's \cite{goldman2008scaling,altshuler2014settling}. The penetration of quasi-2D symmetric disks inside Hele-Shaw cells containing granular matter has been studied both in the vertical \cite{sanchez2014note} and horizontal \cite{kolb2014flow} positions, allowing to quantify the symmetric ``granular flow fields''  surrounding the penetrating object. The ``stopping force'' caused by granular matter has been eventually examined on further symmetrical intruders such as spheres and cones\cite{brzinski2013depth}. Even the ``cooperative'' penetration of several symmetric bodies such as cylinders and disks have been examined both experimentally and through computer simulations. \cite{pacheconc,carvalho2022,Espinosa2022}. Rarely, a non-symmetric scenario has been explored: the sedimentation of a symmetric object into a granular bed, {\it near a single vertical wall} \cite{nelson2008projectile,Diaz-Melian2020,Espinosa2022}: the resulting asymmetry in the force chains developing inside the granular material provokes the repulsion of cylindrical intruders by the wall, and makes them rotate.

More recently, we have examined the effect of asymmetries --or, more generally, imperfections-- {\it in the penetrating object itself}. We have shown that a relatively small perturbation in the shape symmetry or mass distribution of a granular intruder can produce an unexpectedly large effect during sedimentation, such as rotation and, eventually, deviation from the vertical direction \cite{espinosa2023imperfect}. We started observing that, if a rock was released into an air-fluidized granular bed, it consistently tilted in a preferential direction while moving down. A stronger rotation was observed in a ``Janus intruder''. i.e., a cylinder with one of its halves rougher than the other one. In order to quantify the effect of a much smaller irregularity in the trajectory of a sinking object, we systematically studied the penetration of a cylinder with a single row of grains attached to one of its border, into a light granular material. We demonstrated that the irregularity ``concentrates'' chain-like granular forces, producing a net torque, which results in an unexpectedly large rotation of the sinking object. In addition, we theoretically showed that even an non-homogeneous density of of the intruder provokes the tilting effect.

\begin{figure}[!htt]
\includegraphics[width=240px]{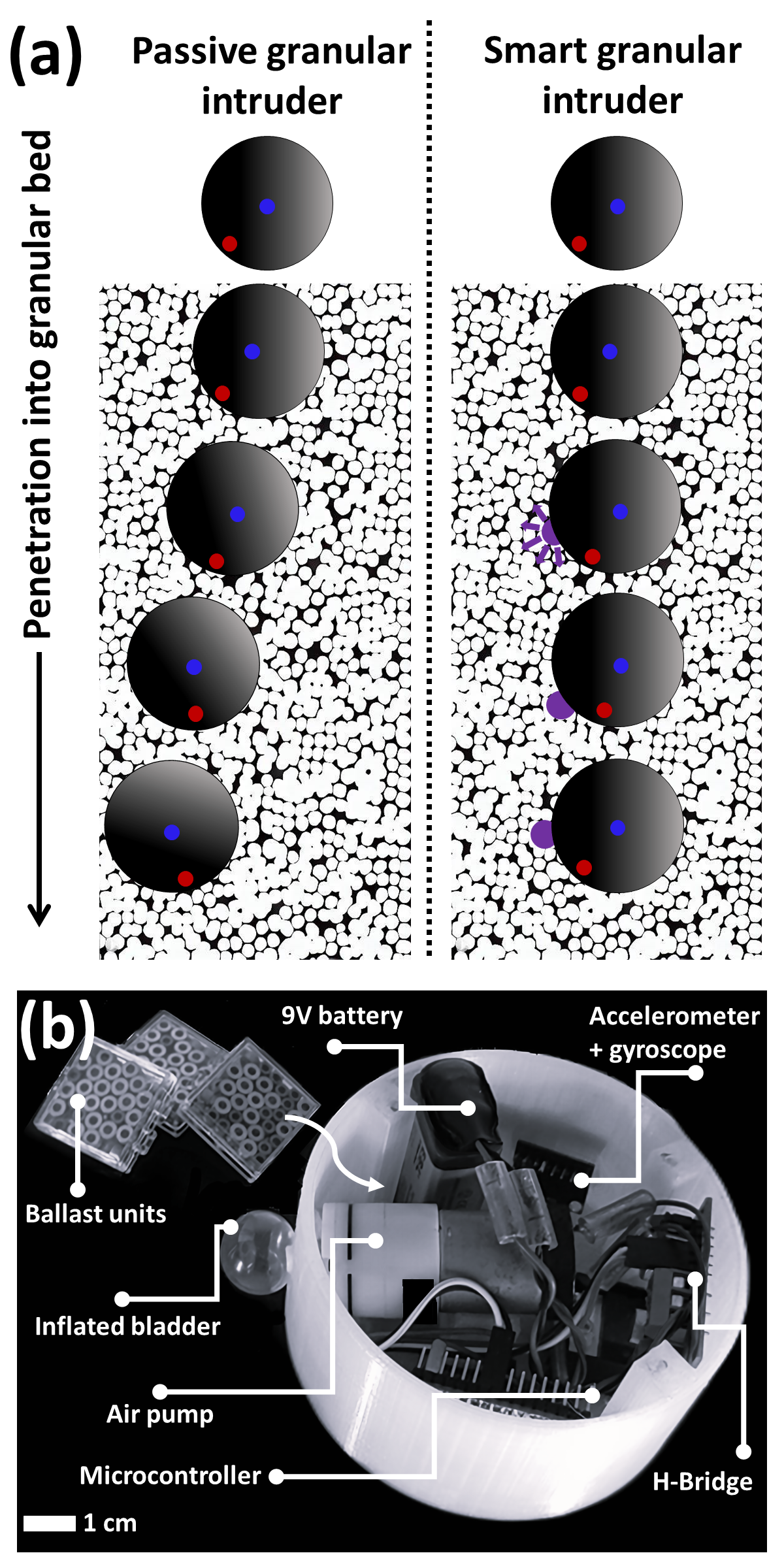}
\caption{Action and anatomy of a smart granular intruder. (a) Sketch of the penetration of a passive (left) and smart (right) granular intruders into a granular bed. From top to bottom, the sketches represent the penetration into the granular bed starting at the free surface, as time goes by. The rotation and lateral deviation of the passive intruder may be provoked by non-homogeneous density, that is controlled by adding ballast units, as shown above. The effect of this asymmetric density is rectified if the intruder becomes ``active''.  (b) Photograph of an activated smart granular intruder with the front lid removed, so the internal components are revealed.}
\label{fig:FigA}
\end{figure}

Here, the basic knowledge discussed above is pulled into a new direction. We propose the idea of mitigating undesirable effects in the settling of granular intruders by equipping them with sensors linked to a soft robotic actuator able to autonomously rectify the sink dynamics. In contrast with previous hard and soft robots designed to walk or swim in sand \cite{li2009sensitive,maladen201undulatory,zhang2013ground,marvi2014sidewinding,valdes2019self,ortiz2019soft}, we aim at solving a specific sedimentation problem in a minimalist way. As rotation is detected by an accelerometer-gyroscope pair inside the intruder, a soft bladder is ordered to inflate and protrude from its surface: the symmetry is then manipulated in such a way that the rotation is minimized. Furthermore, we show that the method not only allows compensating the effects of an intrinsic defect of the intruder, but even path deformations and rotation associated to an external element, such as the presence of a nearby vertical wall. The concept may be scaled up to mitigate the inclination of constructions when settling on partially fluidized soil, and the possible tilting of rovers moving on granular terrain.

\section{METHODS}
\subsection{Design Details of the smart granular intruder}

Our granular intruder consists in a 3D-printed cylinder of $5$ cm radius and $5.2$ cm height, 
that fits between the rear and front glass plates of a Hele-Shaw cell separated by a distance of $5.3$ cm,
that can be partially filled with granular matter (the friction between the circular walls of the intruder and the glass plates was determined to be negligible). Depending on the tilting of the cylinder --measured by appropriate sensors inside it--
it can become ``active'', i.e., a rubber bladder is inflated, protruding from the lateral surface of the cylinder. The bladder strongly modifies the
penetration process, which is sketched in Fig. \ref{fig:FigA}(a). Fig. \ref{fig:FigA}(b) shows a photograph of the real intruder. Now, we explain in more detail the design of the penetrating device.

The cylindrical casing of the intruder was designed in openSCAD and 3D-printed using PETG (modified glycol terephthalate polyethylene) by a Creality Ender 5 Plus printer, with a layer resolution of 0.15 mm. As pointed out in Fig. \ref{fig:FigA} (b), there is a cavity where ballast of three different masses could be positioned, in order to create counter-clockwise rotation (when ballast-free, the device rotates nearly 10 degrees clockwise  when sinking, due to a slight shift of its center of mass, but this has no impact in the outcome of our experiments). The bladder consists in a commercial VIP latex film (Henan Xibei Latex Co., Ltd) attached as a drum-head to the end of a plastic tube of 4 mm inner diameter and 6 mm outer diameter. The tube is connected to a pneumatic pump (KPM27C-6B1 DC, 6V, mini diaphragm pump) able to deliver air at 1.9 liters per minute when turned ON. The rotation angle of the intruder is determined through the utilization of a six-axis MPU-6050 sensor equipped with accelerometers and gyroscopes. Serving as the central processing unit of the robot is the ESP32-WROOM-32 micro-controller, responsible for the management of all internal logic. The power is delivered to the pump by Pulse Width Modulation using a L298N driver, that includes two H-bridge channels able to modulate the power provided by an rechargeable lithium-ion battery (EBL 9V, 600 mAh). The micro-controller communicates with a lab-based PC by HTTP protocol. The intruder total mass (including the internal electronics) is $220.4$ g.

\subsection{Preparation of the granular bed and video acquisition details}

Expanded polystyrene particles used as grains had a density of $0.014\pm0.002$\,g/cc (the intruder/grain density ratio is 96.36) and diameter distributed between $2.0$ and $6.5$\,mm, peaking at $5.8$\,mm. They were deposited into a Hele-Shaw cell of width $55$\,cm and thickness $5.3$\,cm, filled to a height of $40$\,cm \cite{Diaz-Melian2020,Espinosa2022,espinosa2023imperfect}. The intruder could be released from the surface of the granular bed, far from the walls of the cell by means of an electromagnetic device that minimized spurious vibrations and torques on the intruder when released. Before each repetition of the experiment, the robot's gyro sensor was recalibrated to zero. During the course of the experiments and calibrations, three different batteries were kept charging, and we replaced the intruder's battery every three repetitions to ensure that the voltage remained within the appropriate range (see Fig. 4S(a) in Supplemental Material). Two colored dots situated at the center and near the border of one circular face of the cylinder  (see Fig. \ref{fig:FigA}) served as reference points for image analysis. Using them, the motion of the intruder's center of mass could be tracked within an uncertainty of  $0.16$\,mm, and the angle of rotation of the intruder around its symmetry axis could be measured within an uncertainty of $1.03$\,deg, \cite{yupi}. Between experiments, the granular material was removed from the cell, and then it was refilled using a precise protocol: by means of a specially designed funnel with a rectangular cross-section of $2.5\times19.5$\,cm, the granular material was gently deposited from the bottom to a height of approximately $40$\,cm inside the cell, as the funnel was slowly elevated. This resulted in a packing fraction of $0.65 \pm 0.01$. The penetration process was followed using a camera Sony RX100IV with a resolution of $1920$ px$\times$ $1080$ px\, at $250$ frames per second at f=1.8, that took videos through one of the glass faces, over a maximum horizontal  length of 40 cm (angular field of view of 50 degrees), resulting in a minimal spatial resolution of 6.4 $pixels/mm$.

\subsection{DEM simulations}

We performed our numerical simulations using the discrete element model LAMMPS (Large-scale Atomic/Molecular Massively Parallel Simulator) \cite{plimpton2007lammps} to describe and understand the sinking of intruders into a granular bed composed of spherical particles. The intruders were modeled using a $1$~mm-spacing simple cubic lattice, formed by spherical particles with $1$~mm diameter, so they can be assumed as 2D disks.

The key challenge of the method is to numerically solve equations of motion for each particle in the system, which involve the rotation and translation of each element (Equations \ref{e1} and \ref{e2}). In these equations, $i$ represents the index of a particle in the system, $m_i$ is its mass, $\vec{r}_i$ is its position, $\vec{F}_i$ is the net force acting on it, $I_i$ is its moment of inertia, $\vec{\omega}(i)$ is its angular velocity around its own axis, and $\vec{\tau}(i)$ is the total torque applied to particle $i$.

\begin{equation}
m_{i} \dfrac{d^{2}\vec{r}_{i}}{dt^{2}}=\vec{F}_{i}+m_{i}\vec{g}
\label{e1}
\end{equation}
\begin{equation}
I_{i}\dfrac{d \vec{\omega}_{i}}{dt}=\vec{\tau}_{i}
\label{e2}
\end{equation}

The interaction between particles was ruled by a Hertzian contact model \cite{h} using the following parameters: Elastic constant for normal contact $K_n = 3.6 \times 10^5$, elastic constant for tangential contact $K_t = 5.5 \times 10^5$, viscoelastic damping constant for normal contact $\gamma_n = 6.8 \times 10^6$ and viscoelastic damping constant for tangential contact $\gamma_t=\gamma_n/2$. The constants were calculated using the material properties: Young’s Modulus $E = 5$~GPa, coefficient of restitution $e =0.1$, Poisson’s ratio $\nu = 0.28$ and tangential friction $\mu = 0.5$. The effect of these numerical values in the penetration dynamics are discussed in \cite{mes}. In the model, it is assumed that two imaginary springs are associated with the overlap between particles, one in the normal direction and another in the tangential direction at the point of contact. The relative movement of the particles in both directions is related to the harmonic motion of the springs.

The net force $\vec{F}_i$ and the total torque $\vec{\tau}_i$ on a particle $i$ can be calculated using the following equations:
\begin{equation}
\vec{F}_{i} = \sum_{j=1, j \neq i}^{N} (\vec{F}_{n,ij} - \vec{F}_{t,ij} )
\label{e3}
\end{equation}

\begin{equation}
\vec{\tau}_{i} = \sum_{j=1, j \neq i}^{N} (\vec{R}_{ij} \times \vec{F}_{t,ij})
\label{e4}
\end{equation}

Where $\vec{R}_{ij}$ is the vector from the center of particle $i$ to the point of contact with particle $j$,
$\vec{F}_{n,ij}$ and $\vec{F}_{t,ij}$ are the normal and tangential components of the force $\vec{F}_{ij}$, respectively.

The contact force $\vec{F}_{ij}$ between two particles is decomposed into a conservative component ($\vec{F}_{ij}^C$) and
a dissipative component ($\vec{F}_{ij}^D$):
	
	\begin{equation}
\vec{F}_{n,ij}^C = -k_n \delta_n \mathbf{n}_{ij}
\end{equation}

\begin{equation}
\vec{F}_{t,ij}^C = - k_t \delta_t \mathbf{t}_{ij}
\end{equation}

\begin{equation}
\vec{F}_{n,ij}^D = -\eta_n \mathbf{v}_{n,ij} 
\end{equation}

\begin{equation}
\vec{F}_{t,ij}^D = - \eta_t \mathbf{v}_{t,ij}
\end{equation}

Where $k_n$ and $k_t$ are the stiffness coefficients of the model springs in the normal and tangential orientations, respectively. $\delta_n$ and $\delta_t$ are the normal and tangential overlaps between the particles at the point of contact. $\mathbf{n}_{ij}$ and $\mathbf{t}_{ij}$ are the unit vectors in the normal and tangential directions to the surfaces at the point of contact. $\eta_n$ and $\eta_t$ are the normal and tangential damping coefficients, respectively. $\mathbf{v}_{n,ij}$ and $\mathbf{v}_{t,ij}$ are the normal and tangential components of the relative velocity between the particles.

\begin{figure*}[!ht]
\centering
\includegraphics[width=500px]{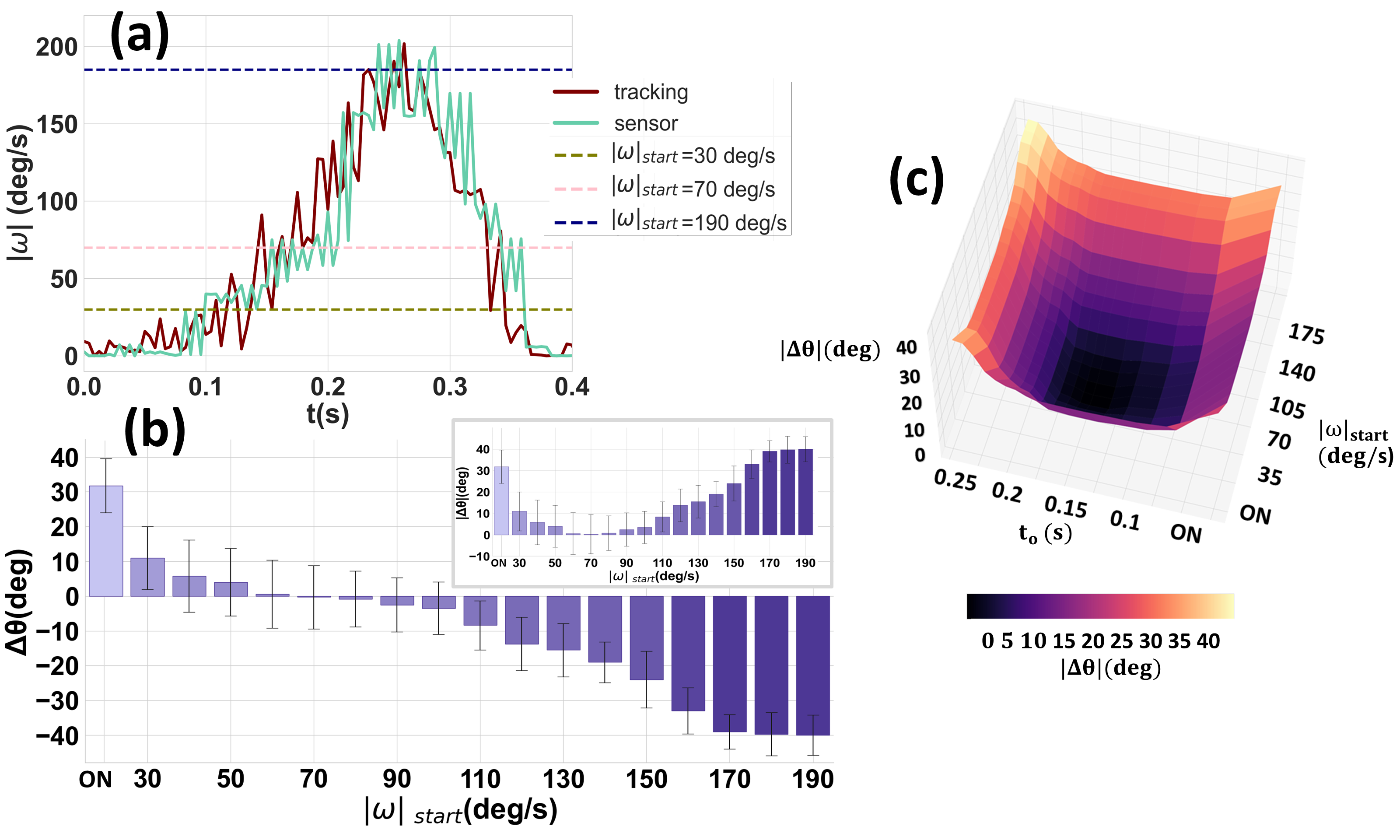}
\centering
\caption{Optimizing the smart granular intruder. (a) Angular velocity of the non-activated intruder as it penetrates a granular bed, measured by the internal sensor and by image analysis: the dotted lines represent the three criteria ($|\omega|_{start}$) used to trigger the pneumatic bladder. (b) Total rotation of the intruder as a function of $|\omega|_{start}$ (the inset shows the module of the rotation). (c) Total rotation of the intruder as a function of both $|\omega|_{start}$ and the time where the pneumatic pump is activated, $t_0$. In (b), each bar represents averaged values over $10$ repetitions of the experiment, and the black segments represents the standard deviation. All data correspond to an intruder with added mass of $2\times 21.64$~g.}
\label{fig:FigB}
\end{figure*}

We initiated each simulation by preparing the granular bed, pouring batches of particles with radius following a uniform random distribution between $1 - 3.25$~mm and a fixed density of $14\times10^{-3}$~g/cc. Each pour generated particles at random positions in a limited space of the container (cuboid with the same experimental dimensions) that moved in the $z$-axis as the container was filled, resulting in a total of $10^5$ particles. The intruders were released from the granular surface with zero initial velocity after the system relaxes. The protocol used in the simulations to activate the bladder is the same as that used in the experiments (see Fig. 6S in Supplemental Material). When $|\omega|$ reaches 70 deg/s, a grain with a initial radius of 0.1 mm begins to grow at a rate of 14.3676 cm$^3$/s. The packing fraction of the granular bed used in the simulations was similar to the one measured in the experiments.

\section{Results and discussion}

Fig. \ref{fig:FigA}(a) sketches a sequence of snapshots representing the penetration of a cylindrical intruder into a granular bed contained in a Hele-Shaw cell: time and penetration depth increases from top to bottom (the top picture represents the intruders slightly above the free granular surface). A passive intruder with a mass asymmetry (in our case, the left half of the cylinder is heavier than the right one, which is represented by darker-lighter gray) penetrates and rotates counter-clockwise, as represented in the left column \cite{espinosa2023imperfect}. The right column sketches the penetration of a smart granular intruder (SGI). Immediately after being released, the SGI rotates counter-clockwise, just as a passive intruder with a mass asymmetry. However, as soon as its angular velocity matches a given threshold ($|\omega|_{start}$) --which is detected by an accelerometer+gyroscope sensor inside-- a micro-controller orders an air pump to inflate a soft bladder, represented in violet in the sketch. In spite of its relatively small size, the resulting bubble-like protrusion exerts a strong effect on the penetration dynamics: the counterclockwise rotation eventually stops, and can be even compensated by clockwise rotation, in such a way that the time-averaged rotation of the SGI becomes negligible. An advantage of using an angular velocity threshold to inflate the bladder is that it always grows against grains that are already ``mobilized'' around the intruder, so the ``local'' packing fraction is not too high, making easier the inflation process.

\begin{figure*}[!t]
\centering
\includegraphics[width=300px]{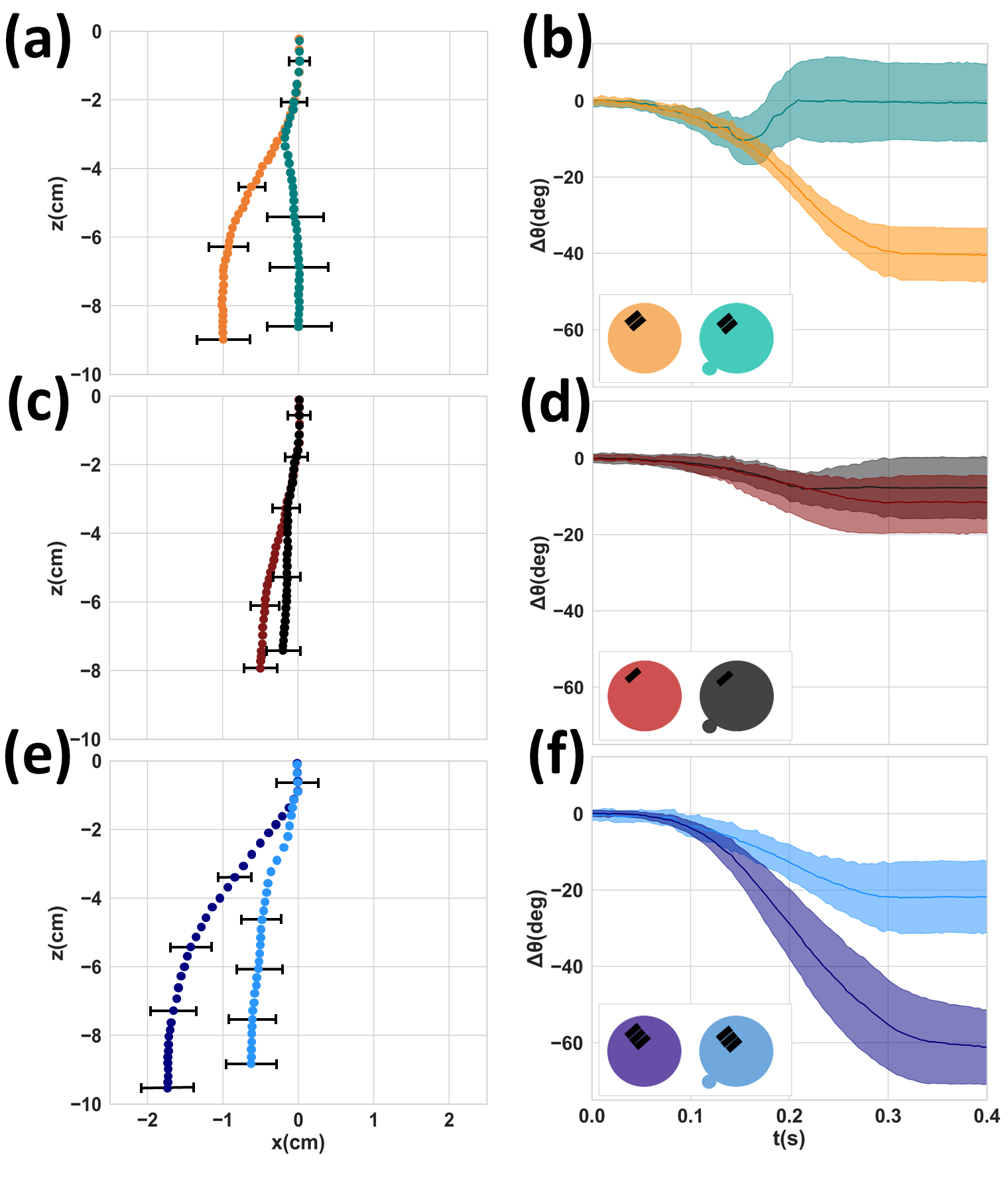}
\centering
\caption{Smart intruder dynamics: (a), (c) and (e) Trajectories of the intruder with and without activating the pneumatic bladder for added masses of $2 \times 21.64$~g, $21.64$~g and $3 \times 21.64$~g, respectively. (b), (d) and (f) Rotation of the intruder with and without activating the pneumatic bladder for added masses of $2 \times 21.64$~g, $21.64$~g and $3 \times 21.64$~g, respectively. In all cases, curves represents averaged values over $10$ repetitions of the experiment, the black bars and the colored widths represent the corresponding standard deviations.}.
\label{fig:FigC}
\end{figure*}

Fig. \ref{fig:FigA}(b) shows a photograph of an actual SGI with the front lid removed so its internal components can be visualized. It consists in a $10$-cm diameter, $5.2$-cm height 3D printed cylinder, with a total weight of $220.4$~g when the lid is re-installed. Besides the elements mentioned above, the photograph show ``ballast units'' of $21.64$~g each, that can be inserted into the intruder to increase the mass of their left side. In the photograph, the intruder is in the ``active'' mode, i.e., the pneumatic bladder is inflated.

Let us assume that we have prepared an intruder with a given mass asymmetry, thanks to the addition of two ballast units in the position illustrated in Fig. \ref{fig:FigA}(b). It is then released into a granular bed deposited inside a Hele-Shaw cell following the protocol described in METHODS. If the intruder is released from the free granular surface without activating the bladder over the entire penetration process, it will penetrate vertically a distance $z_{max}$ and will rotate counterclockwise a total angle of approximately $40^o$ until the motion stops \cite{espinosa2023imperfect}. As described earlier, the aim of the SGI is to minimise the undesired rotation by autonomously activating the pneumatic bladder in the right moment. We have chosen to use the angular velocity of the intruder around its symmetry axis as the critical parameter to activate the bladder's inflation: as stated above, if $|\omega| < |\omega|_{start}$ the air pump is off; if $|\omega| \geq |\omega|_{start}$, the air pump is turned on, so the bladder inflates to a final diameter of approximately $1.4$~cm in nearly $0.1$~s (see Fig. 4S in Supplemental Material for further details).

Fig. \ref{fig:FigB}(a) shows the temporal evolution the angular rotation speed as time goes by for a non-activated smart granular intruder, demonstrating the excellent match between a measurement based on the internal sensor, and on external image analysis. The horizontal discontinuous lines illustrate three different $|\omega|_{start}$ values used to trigger the inflation of the bladder.

Fig. \ref{fig:FigB}(b) displays the total rotation angle of the intruder as a function of several $|\omega|_{start}$ values, including the three ones indicated in Fig. \ref{fig:FigB}(a), as well as the case were the bladder is inflated from the beginning of the experiment (``ON''). The graph reveals that there is an almost horizontal plateau of $|\omega|_{start}$ values between $60$ and $80$~deg$/$s where the action of the bladder almost totally rectifies the average rotation. This demonstrates that our SGI is able to compensate the undesired rotation, and that $|\omega|_{start}$ is a simple, yet effective parameter to activate the bladder. Later on, we will study in more detail the penetration of a SGI operating with an ``optimal'' threshold located at the centre of the plateau, i.e., $|\omega|_{start}^{optimal}=70$~deg$/$~s. It must be stressed that the choice of $\omega_{start}$ has also a moderate influence on the total vertical penetration (not studied here), which can be quite important is some potential applications.

Fig \ref{fig:FigB}(c) shows how the total rotation angle depends on both $|\omega|_{start}$ and the time  at which the bladder starts to inflate, $t_0$: it exhibits a  minimum centered at $|\omega|_{start}^{optimal}$ and $t_0 \approx 0.15$~s. ($t_0$ was determined by an acoustic method described in detail in the Supplemental Material).

We now delineate the dynamic progression of the penetration process across varying mass distributions, both in the presence and absence of inflatable bladder activation. Figure \ref{fig:FigC} illustrates the trajectories of the intruder (left column) and the temporal evolution of rotational behavior (right column). for $|\omega|_{start}^{optimal}=70$~deg$/s$. The first row corresponds to an added mass of $2 \times 21.64$~g (illustrated as two black bars inside the intruder), which was the one used for optimizing $|\omega|_{start}$, as mentioned earlier. Without activating the bladder, the net lateral motion of the center of mass is of approximately $1$~cm to the left in average (Fig. \ref{fig:FigC}(a)), while the net rotation reaches $40$~degrees counterclockwise (Fig. \ref{fig:FigC}(b)). When the bladder is activated at $|\omega|_{start}^{optimal}=70$~deg$/s$, the compensation effect starts at $t_0 \approx 0.15$~s, almost fully suppressing undesired motions at $t \approx 0.3$~s for $\Delta x$, and as early as $t \approx 0.2$~s for $\Delta \theta$. So, our protocol is surprisingly robust for early rectification of ``spurious'' rotation and lateral migration.

The second row corresponds to an added mass of $21.64$~g (one black bar inside the intruder), producing a lateral motion of $0.5$~cm to the left (Fig. \ref{fig:FigC} (c)) and a net counterclockwise rotation angle in excess of $15$~degrees (Fig. \ref{fig:FigC}(d)) without bladder action. As the bladder is activated, both the lateral motion and the rotation are partially compensated: the net horizontal displacement is now of approximately $0.25$~cm to the left (Fig. \ref{fig:FigC}(c)), while the net rotation is of nearly $8$~degrees counterclockwise (Fig. \ref{fig:FigC}(d)). The third row corresponds to an added mass of $3 \times 21.64$~g (three black bars inside the intruder), producing a lateral motion of nearly $1.8$~cm to the left (Fig. \ref{fig:FigC}(e)) and a net counterclockwise rotation in excess of $60$~degrees (Fig. \ref{fig:FigC}(f)) without bladder action. When the bladder is activated, both the lateral motion and the rotation are also partially compensated: the net horizontal displacement is somewhat larger than $0.5$~cm to the left (Fig. \ref{fig:FigC}(e)), and the net rotation is of approximately $20$~degrees counterclockwise. While these values may look large, the bladder is able to reduce both $\Delta x$ and $\Delta \theta$ to nearly one-third of their net values in the absence of bladder activation. Notice that the results illustrated in Fig. \ref{fig:FigC}(c) - (f) were obtained for $|\omega|_{start}$ {\it optimized for a different intruder}. These results suggest that, while the bladder action always tends to rectify the spurious motion to a non-negligible degree, it is important to optimize it for the specific kind of intruder.

\begin{figure}[!ht]
\includegraphics[width=250px]{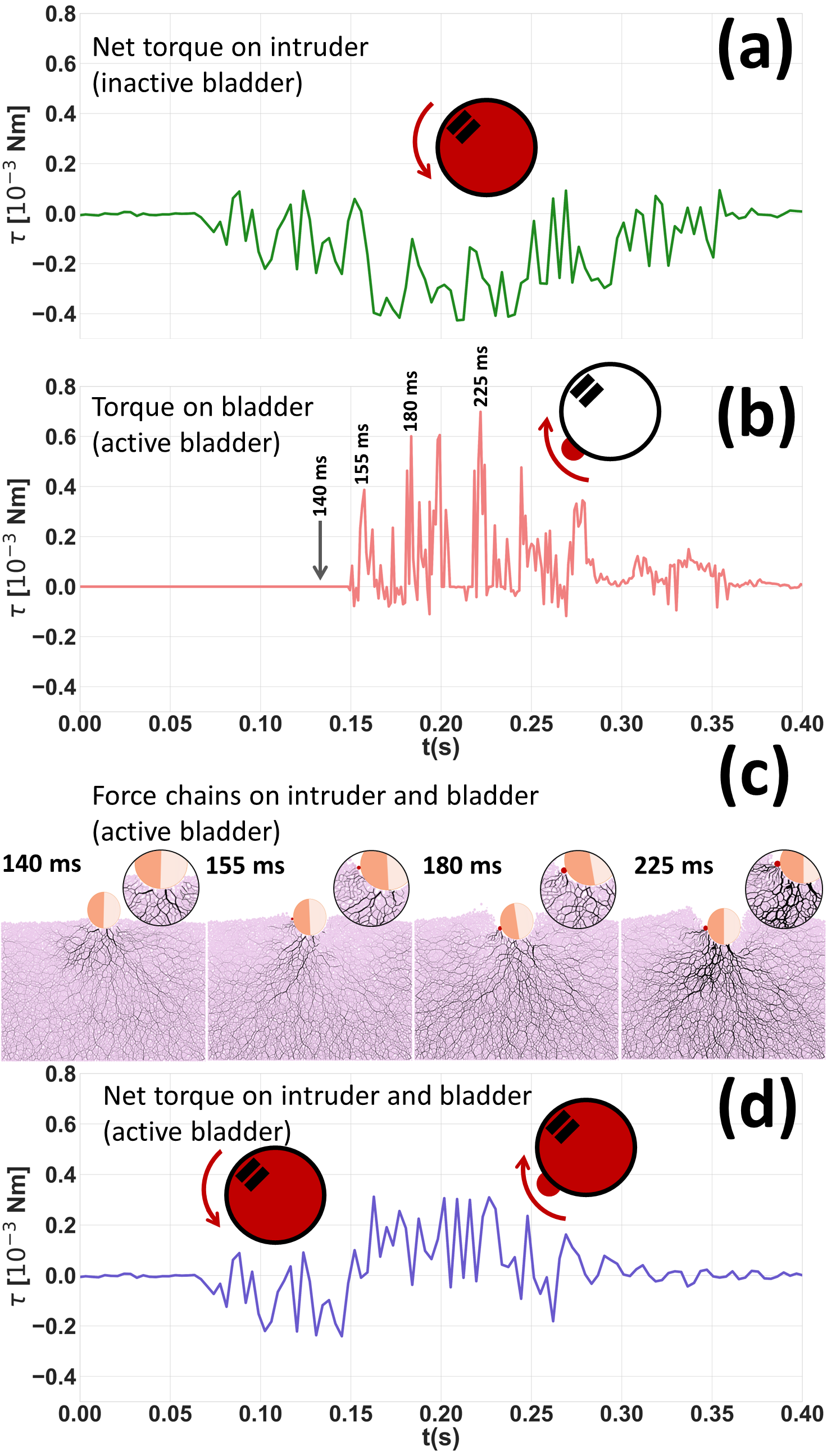}
\centering
\caption{Smart intruder's mechanism revealed by DEM simulations. (a) Net torque on the intruder with inactivated bladder as it penetrates the granular medium. (b) Torque felt by the inflated bladder. (c) Granular force chains associated to the whole intruder (including activated bladder) as time goes by. The insets correspond to zooms in the region near the bladder. (d) Net torque on the intruder (including activated bladder). Negative and positive torques correspond to counterclockwise and clockwise motion, respectively.}
\label{fig:FigD}
\end{figure}

In order to shed light on the role of the activated bladder, we made DEM simulations of a mass-imbalanced disk without and with a single grain attached to its border that changes its size at the same rate as the real bladder does (for more details on the simulations see Supplemental Material). Fig. \ref{fig:FigD}(a) shows the net torque on the mass-imbalanced intruder without the effect of the bladder. As expected, the excess internal mass at the left of the disk produces a torque responsible for the anti-clockwise rotation, that we have defined as negative; the ``spiky'' nature of the torque will be discussed below. Fig. \ref{fig:FigD}(b) shows the torque produced by the grains exclusively on the bladder, which is now activated. Here, the action is even more ``spiky'', but always positive: it ``tries to compensate'' the overall torque in the absence of the inflated bladder (the larger ``spikiness'' of the torques on the bladder might be related to the importance of fluctuations acting on a smaller object, in contrast to the whole intruder). The bottom panel in Fig. \ref{fig:FigD}(c) reveals the cause of the torque spikes seen in (b) by representing the granular force chains at four different moments of the penetration. At $140$~ms the bladder is still not inflated, so all the force chains are more or less symmetrically applied at the bottom half of the disk. At $155$, $180$ and $225$~ms there are extra, increasing concentrations of force chains at the bladder (corresponding to strong spikes in (b)) on top of the roughly symmetrical force chains acting on the bottom half of the disk. This ``symmetry breaking'' of the overall distribution of force chains due to the presence of the inflated bladder provokes the net result visualised in Fig. \ref{fig:FigD}(d): in the first half of the penetration process, the net torque is negative, but in the second half (starting at $t_0 \approx 0.15$~s) the torque associated to the inflated bladder takes over, so the net torque becomes positive. The result is a ``rectification'' like that illustrated in Fig. \ref{fig:FigC}(a) and (b).

Finally, we have also performed preliminary experiments demonstrating that the smart intruder concept can be used to compensate the rotation of a cylindrical object produced by an {\it external} asymmetry. In particular, we have experimentally shown that a pneumatic bladder inflating on the side of the cylinder opposite to a nearby boundary strongly compensates the ``rolling away from the wall'' effect due to the proximity of the boundary \cite{Diaz-Melian2020} (see Supplemental Material).

\section{Conclusion}
We have designed and constructed a cylindrical granular intruder with a soft, pneumatic bladder that can robustly compensate undesired rotations caused by either external causes or intrinsic asymmetries of the object. The smart intruder operates autonomously, based on the input from internal sensors: if its rotation speed is larger than a given threshold as measured by internal sensors, the bladder is activated, suppressing the undesired rotation. Moreover, we have shown by discrete elements simulations that the smart intruder's mechanism is based on the fact that granular force chains concentrate on the bladder, resulting in a torque that compensates the opposite torque causing the undesired rotation. We believe that this general concept can be extrapolated to create active building foundations reacting to partial soil fluidization, thus mitigating structural tilting. However, realistic systems involving many degrees of freedom may require an extended version of the mechanism we have proposed here.

\section*{Supplementary Material}
A PDF containing Figures 1S to 7S. This document provides additional information regarding the design and operation of the SGI. It also includes details about DEM simulations and other experimental results.

\section*{ACKNOWLEDGMENTS}
L. A. Rodr\'{\i}guez-de-Torner are gratefully acknowledged for helping with the pneumatic design of the intruder, and A. Petrof for providing support for video recording. E. Altshuler found inspiration in the late M. \'Alvarez-Ponte. This work is not funded by any grants. 
\section*{AUTHOR DECLARATIONS}
\subsection*{Conflict of Interest}
The authors have no conflicts to disclose.

\subsection*{Author Contributions} E. A. designed research. L.M.-O., E. A. and A. R-R designed and constructed the smart intruder.  L.M.-O. conducted the experiments, extracted, and processed the data. L.M.-O. performed the simulations. L.M.-O and E.A. interpreted results. E.A. and L.M.-O. wrote the article. 
\section*{DATA AVAILABILITY}
The data that support the findings of this study are available from
the corresponding authors upon reasonable request.


\section*{References}
\begin{thebibliography}{4}
\bibitem{sancio2003ground}
R.~B. Sancio, 
\newblock {``Ground failure and building performance in Adapazari, Turkey''},
\newblock University of California, Berkeley,  (2003).

\bibitem{kirca2019sinking}
V. Kirca, and B.~M. Sumer,
\newblock ``Sinking of Anchors and Other Subsea Structures due to Wave-Induced
  Seabed Liquefaction'',
\newblock {\em Coastal Structures 2019}{ \bf }, 598--607 (2019).

\bibitem{cai2021experimental}
 Y. Cai, Q.-f. Liu, L. Yu, Z. Meng, Z. Hu, Q Yuan, and B. {\v{S}}avija, 
\newblock ``An experimental and numerical investigation of coarse aggregate
  settlement in fresh concrete under vibration'',
\newblock {\em Cement and Concrete Composites}{ 122}, 104153 (2021).

\bibitem{Mickleburgh2018actualistic}
H.~L. Mickleburgh, 
\newblock ``Actualistic experimental taphonomy of inhumation burial'',
\newblock In {\em Multidisciplinary approaches to forensic archeology},
  105--114. Springer (2018).

\bibitem{Moeyersons1977}
J. Moeyersons,
\newblock ``The behaviour of stones and stone implements, buried in consolidating and creeping Kalahari sands'',
\newblock {\em Earth Surf. Processes}{ 3}, 115 (1977).

\bibitem{RoverSpirit2010}
Jaggard, Victoria: 
\newblock Mars Rover to Roam No More -- It's Official,
  (https://www.nationalgeographic.com/science/article
  \\/100126-mars-rover-spirit-nasa-stuck).








\bibitem{katsuragi2007unified}
H. Katsuragi, and D.~J. Durian,
\newblock ``Unified force law for granular impact cratering'',
\newblock {\em Nat. Phys.}{ 3}(6), 420 (2007).

\bibitem{pacheco2011infinite}
F. Pacheco-V{\'a}zquez, G. Caballero-Robledo, J. Solano-Altamirano, E. Altshuler, A. Batista-Leyva, and J. Ruiz-Su{\'a}rez,
\newblock ``Infinite penetration of a projectile into a granular medium'',
\newblock {\em Phys. Rev. Lett.}{ 106}(21), 218001 (2011).


\bibitem{goldman2008scaling}
D.~I. Goldman, and P. Umbanhowar,
\newblock ``Scaling and dynamics of sphere and disk impact into granular media'',
\newblock {\em Phys. Rev. E}{ 77}(2), 021308 (2008).

\bibitem{altshuler2014settling}
E. Altshuler, H. Torres, A. Gonz{\'a}lez-Pita, G. S{\'a}nchez-Colina, C. P{\'e}rez-Penichet, S. Waitukaitis, and R. Hidalgo,
\newblock ``Settling into dry granular media in different gravities'',
\newblock {\em Geophys. Res. Lett.}{ 41}(9), 3032--3037 (2014).

\bibitem{sanchez2014note}
G. S{\'a}nchez-Colina, L. Alonso-Llanes, E. Mart{\'\i}nez, A. Batista-Leyva, C. Clement, C. Fliedner, R. Toussaint, and E. Altshuler,
\newblock ``Note:“Lock-in accelerometry” to follow sink dynamics in shaken
  granular matter'',
\newblock {\em Rev. Sci. Instrum.}{ 85}(12), 126101 (2014).

\bibitem{kolb2014flow}
E. Kolb, P. Cixous, and J. Charmet,
\newblock ``Flow fields around an intruder immersed in a 2D dense granular layer'',
\newblock {\em Granular Matter}{ 16}(2), 223--233 (2014).

\bibitem{brzinski2013depth}
T.~A. Brzinski~III, P. Mayor, and D.~J. Durian,
\newblock ``Depth-dependent resistance of granular media to vertical penetration'',
\newblock {\em Phys. Rev. Lett.}{ 111}(16), 168002 (2013).
https://www.overleaf.com/project/65b08fa070ffb3c7dbcbb895

\bibitem{pacheconc}
F. Pacheco-Vázquez, J. Ruiz-Suárez,  
\newblock ``Cooperative dynamics in the penetration of a group of intruders in a granular medium'', \newblock {\em Nature communications} 1.1 (2010): 123.

\bibitem{carvalho2022}
D. D. Carvalho, and E. M. Franklin, 
\newblock ``Collaborative behavior of intruders moving amid grains'', 
\newblock {\em Physics of Fluids} 34.12 (2022).

\bibitem{Espinosa2022}
M. Espinosa, V. Diaz-Meli\'{a}n, A. Serrano-Mu\~{n}oz, and E.Altshuler,
\newblock ``Intruders cooperatively interact with a wall into granular matter'',
\newblock {\em Granular Matter}{ 24}(39), 2 (2022).

\bibitem{nelson2008projectile}
E. Nelson, H. Katsuragi, P. Mayor, and D.~J. Durian,
\newblock ``Projectile interactions in granular impact cratering'',
\newblock {\em Phys. Rev. Lett.}{101}(6), 068001 (2008).


\bibitem{Diaz-Melian2020}
V. Diaz-Meli\'{a}n, A. Serrano-Mu\~{n}oz, M. Espinosa, L. Alonso-Llanes,
  G. Viera-Lopez, and E. Altshuler,
\newblock ``Rolling away from the wall into granular matter'',
\newblock {\em Phys. Rev. Lett.}{ 125}(1), 078002 (2020).



\bibitem{espinosa2023imperfect}
M. Espinosa, L. Mart{\'\i}nez-Ort{\'\i}z, L. Alonso-Llanes, L. A. Rodr{\'\i}guez-de-Torner, O. Ch{\'a}vez-Linares, and E. Altshuler, 
\newblock ``Imperfect bodies sink imperfectly when settling in granular matter'',
\newblock {\em Sci. Adv.}{ 9}(19), eadf6243 (2023).

\bibitem{li2009sensitive}
C. Li, P. B Ambanhowar, H. Komsuoglu, D. E. Koditschek, and D. I. Goldman,
\newblock ``Sensitive dependence of the motion of a legged robot on granular media'',
\newblock {\em Proceedings of the National Academy of Sciences}{ 106}(9), 3029--3034 (2009).

\bibitem{maladen201undulatory}
R. D. Maladen, Y. Ding, P. B. Umbanhowar, and D. I. Goldman,
\newblock ``Undulatory swimming in sand: experimental and simulation studies of a robotic sandfish'',
\newblock {\em The International Journal of Robotics Research}{ 30}(7), 793--805 (2011).

\bibitem{zhang2013ground}
T. Zhang, F. Qian, C. Li, P. Masarati, A. M. Hoover, P. Birkmeyer, A. Pullin, R. S. Fearing, and D. I. Goldman,
\newblock ``Ground fluidization promotes rapid running of a lightweight robot'',
\newblock {\em The International Journal of Robotics Research}{ 32}(7), 859--869 (2013).

\bibitem{marvi2014sidewinding}
H. Marvi, Ch. Gong, N. Gravish, H. Astley, M. Travers, R. L. Hatton, J. R. Mendelson III, H. Choset, D. L. Hu, and D. I. Goldman,
\newblock ``Sidewinding with minimal slip: Snake and robot ascent of sandy slopes'',
\newblock {\em Science}{ 346}(6206), 224--229 (2014).

\bibitem{valdes2019self}
R. Vald{\'e}s, V. Angeles, E. de la Calleja, and R. Zenit,
\newblock ``Self-propulsion of a helical swimmer in granular matter'',
\newblock {\em Physical Review Fluids}{ 4}(8), 084302 (2019).

\bibitem{ortiz2019soft}
D. Ortiz, N. Gravish, and M. T. Tolley,
\newblock ``Soft robot actuation strategies for locomotion in granular substrates'',
\newblock {\em IEEE Robotics and Automation Letters}{ 4}(3), 2630--2630 (2019).

\bibitem{yupi}
A. Reyes, G. Viera-López, J.J Morgado-Vega, and E. Altshuler,
\newblock ``yupi: Generation, tracking and analysis of trajectory data in Python'',
\newblock {\em Environmental Modelling $\&$ Software}{ 163}, (2023).

\bibitem{plimpton2007lammps}
S. Plimpton,
\newblock ``Fast parallel algorithms for short-range molecular dynamics'',
\newblock {\em J. Comp. Phys.}{ 117}, 1--19 (1995).



\bibitem{h}
NV Brilliantov, F Spahn, JM Hertzsch, and T P¨oschel.
\newblock Model for collisions in granular gases. 
\newblock {\em Physical review E}, 53(5):5382, (1996)


\bibitem{mes}
M Espinosa, and E Altshuler. 
\newblock Impact of grain properties on the penetration of intruders near a wall into granular matter: a DEM study. 
\newblock {\em Revista Cubana de Física}, 39, 8 (2022).

\end{thebibliography}
\end{document}